\documentclass[aps,prb,twocolumn,superscriptaddress]{revtex4-1}
\usepackage[colorlinks=true,citecolor=blue,linkcolor=blue,urlcolor=blue]{hyperref}
\usepackage{graphicx, nicefrac}
\usepackage{hyperref}

\begin{document}

\title{Resonant inelastic x-ray scattering study of bond order and spin excitations in nickelate thin-film structures}

\author{K. F{\"u}rsich}
\author{Y. Lu}
\affiliation{Max-Planck-Institut f{\"u}r Festk{\"o}rperforschung, Heisenbergstrasse 1, 70569 Stuttgart, Germany}
\author{D. Betto}
\affiliation{Max-Planck-Institut f{\"u}r Festk{\"o}rperforschung, Heisenbergstrasse 1, 70569 Stuttgart, Germany}
\affiliation{European Synchrotron Radiation Facility, 71 Avenue des Martyrs, Grenoble F-38043, France}
\author{M. Bluschke}
\affiliation{Max-Planck-Institut f{\"u}r Festk{\"o}rperforschung, Heisenbergstrasse 1, 70569 Stuttgart, Germany}
\affiliation{Helmholtz-Zentrum Berlin f{\"u}r Materialien und Energie, Wilhelm-Conrad-R{\"o}ntgen-Campus BESSY II, 12489 Berlin, Germany}
\author{J. Porras}
\affiliation{Max-Planck-Institut f{\"u}r Festk{\"o}rperforschung, Heisenbergstrasse 1, 70569 Stuttgart, Germany}
\author{E. Schierle}
\affiliation{Helmholtz-Zentrum Berlin f{\"u}r Materialien und Energie, Wilhelm-Conrad-R{\"o}ntgen-Campus BESSY II, 12489 Berlin, Germany}
\author{R. Ortiz}
\author{H. Suzuki}
\author{G. Cristiani}
\author{G. Logvenov}
\affiliation{Max-Planck-Institut f{\"u}r Festk{\"o}rperforschung, Heisenbergstrasse 1, 70569 Stuttgart, Germany}
\author{N.B. Brookes}
\affiliation{European Synchrotron Radiation Facility, 71 Avenue des Martyrs, Grenoble F-38043, France}
\author{M.W. Haverkort}
\affiliation{Institut f{\"u}r Theoretische Physik, Universit{\"a}t Heidelberg, Philosophenweg 19, 69120 Heidelberg, Germany}
\author{M. Le Tacon}
\affiliation{Karlsruher Institut f{\"u}r Technologie, Institut f{\"u}r Festk{\"o}rperphysik, Hermann-v.-Helmholtz-Platz 1, 76344 Eggenstein-Leopoldshafen, Germany}
\author{E. Benckiser}
\affiliation{Max-Planck-Institut f{\"u}r Festk{\"o}rperforschung, Heisenbergstrasse 1, 70569 Stuttgart, Germany}
\author{M. Minola}
\email{M.Minola@fkf.mpg.de}
\author{B. Keimer}
\email{B.Keimer@fkf.mpg.de}
\affiliation{Max-Planck-Institut f{\"u}r Festk{\"o}rperforschung, Heisenbergstrasse 1, 70569 Stuttgart, Germany}

\date{\today}

\begin{abstract}
We used high-resolution resonant inelastic x-ray scattering (RIXS) at the Ni $L_3$ edge to simultaneously investigate high-energy interband transitions characteristic of Ni-O bond ordering and low-energy collective excitations of the Ni spins in the rare-earth nickelates $R$NiO$_3$ ($R$ = Nd, Pr, La) with pseudocubic perovskite structure. With the support of calculations based on a double-cluster model we quantify bond order (BO) amplitudes for different thin films and heterostructures and discriminate short-range BO fluctuations from long-range static order. Moreover we investigate magnetic order and exchange interactions in spatially confined $R$NiO$_3$ slabs by probing dispersive magnon excitations. While our study of superlattices (SLs) grown in the (001) direction of the perovskite structure reveals a robust non-collinear spin spiral magnetic order with dispersive magnon excitations that are essentially unperturbed by BO modulations and spatial confinement, we find magnons with flat dispersions and strongly reduced energies in SLs grown in the $(111)_{\text{pc}}$ direction that exhibit collinear magnetic order. These results give insight into the interplay of different collective ordering phenomena in a prototypical 3$d$ transition metal oxide and establish RIXS as a powerful tool to quantitatively study several order parameters and the corresponding collective excitations within one experiment.
\end{abstract}

\pacs{df}

\maketitle

\section{Introduction}
Nickelate perovskites of composition $R$NiO$_3$ (with $R$ = rare earth) are archetypes of correlated-electron behavior in the vicinity of a Mott metal-insulator transition (MIT) \cite{Torrance92, GarciaMunoz92, Medarde97, Guo18}. The microscopic origin of the MIT and of an unusual antiferromagnetic (AFM) state in the insulating phase have stimulated a great deal of theoretical work over several decades \cite{Mizokawa00, Lee11, Lee11a, Park12, Subedi15, Green16, Varignon17}. New perspectives have recently emerged from the ability to synthesize thin-film structures with atomic-scale precision \cite{Hwang12, Catalano15} and to probe ordering phenomena in such structures using Ni $L$ edge resonant x-ray scattering \cite{Frano13, Wu13}. Resonant elastic x-ray scattering from heterostructures with atomically thin nickelate layers revealed magnetic ground states different from those of bulk $R$NiO$_3$ \cite{Frano13, Hepting14}. Very recently, advances in improving the energy resolution of resonant inelastic x-ray scattering (RIXS) have enabled the observation of dispersive spin excitations in nickelate films. The resulting data provide detailed information on the exchange interactions that drive magnetic order in these systems \cite{Lu18}. Here we take advantage of high-resolution RIXS to probe manifestations of Ni-O bond order and fluctuations in high-energy interband (``$dd$'') transitions, and to investigate collective spin excitations in heterostructures exhibiting magnetic ground states different from those in the bulk.\\
All $R$NiO$_3$ with $R\neq\text{La}$ show a MIT at a temperature that decreases monotonically as a function of increasing rare-earth ionic radius, which straightens the Ni-O-Ni bonds and enhances the valence-electron bandwidth. The MIT is accompanied by a structural transition \cite{Zaghrioui01}, where a pattern of alternating NiO$_6$ octahedra volumes develops along the diagonal of the pseudocubic (pc) perovskite structure with the ordering vector $\textbf{Q}_{\text{BO}}=(\nicefrac{1}{2}, \nicefrac{1}{2},\nicefrac{1}{2})$ \cite{Alonso99, Alonso00}. Different models have been proposed to describe the insulating state and $\textbf{Q}_{\text{BO}}$. While early studies ascribed the insulating state to charge disproportionation on the Ni site \cite{Medarde97}, recent experimental and theoretical work has pointed out that $\textbf{Q}_{\text{BO}}$ actually results from ordering of the Ni-O bonds \cite{Johnston14, Park12}. The key to this finding is the \textit{self-doped} ground state of  $R$NiO$_3$, where an electron is transferred from the oxygen ligands to the Ni 3$d$ orbital, effectively resulting in a $d^8\underline{L}^1$ configuration \cite{Mizokawa00}. Here $\underline{L}$ stands for a hole on the oxygen ligands. According to this model, the electrons rearrange on the oxygen ligands at the MIT, whereas the Ni ions remain in the $d^8$ state. Consequently, alternating octahedra with longer (LB) and shorter (SB) Ni-O bond-lengths are formed in the insulating phase, which have a $d^8\underline{L}^0$ $(S\approx1)$ and $d^8\underline{L}^2$ $(S\approx0)$ configuration, respectively \cite{Green16}.\\
In addition to the BO, $R$NiO$_3$ host an unusual AFM order, which either develops simultaneously with the MIT upon cooling (for $R$ = Nd, Pr) or at lower temperatures in the insulating state (for smaller $R$) \cite{Vobornik99}. Early neutron powder diffraction experiments on  $R$NiO$_3$ found the magnetic ordering vector \textbf{Q}$_{\text{AFM}}=(\nicefrac{1}{4}, \nicefrac{1}{4},\nicefrac{1}{4})_{\text{pc}}$ \cite{Garcia94, Rodriguez98}. Two different spin structures were discussed to explain this peculiar ordering vector: a collinear up-up-down-down state or a non-collinear spin spiral. 
Independent magnetic x-ray scattering experiments later demonstrated a spin-spiral magnetic ground state in bulk-like films of PrNiO$_3$ and NdNiO$_3$ \cite{Scagnoli06, Scagnoli08, Frano13}.\\
The spin and bond order in $R$NiO$_3$ can be further tuned by different external parameters, such as pressure, epitaxial strain or reduced dimensionality, therefore providing an excellent playground to study the interplay of the collective ordering phenomena \cite{Frano13, Hepting14, Liu10, Middey16review}. The spatial confinement achieved in heterostructures allows one to selectively tune magnetic and bond order. In particular, one can obtain ground states that do not occur in bulk-like films, such as a metallic state where magnetic order persists in absence of BO, which is of potential interest for spintronic applications \cite{Lu16, Hepting14}.  In heterostructures of insulating nickelates and non-magnetic metal oxides, one can realize a collinear spin structure by suppressing BO and truncating the exchange bonds \cite{Hepting18}.
Along these lines, recent theoretical and experimental studies suggest that the mechanism of the MIT differs for bulk and ultrathin nickelate layers \cite{Lee11, Lu17}.\\
Additionally, some studies indicate a close feedback between AFM and BO, such that the presence of the former profoundly modulates the bond-disproportionation amplitude \cite{Ruppen2017, Hampel17}. To further investigate the interactions and hierarchy of different ordering phenomena in $R$NiO$_3$ and to test corresponding model calculations, it is therefore crucial to experimentally determine both bond and magnetic order quantitatively on the same samples.\\
Single crystalline $R$NiO$_3$ are available to date only as thin layers in films and heterostructures. Since the scattering volume  is too small for neutrons, momentum-resolved experiments to access AFM and BO can be carried out only using x-rays. \\
RIXS has been proven to be an excellent tool to study the electronic properties of correlated oxides.
Specifically, the recent  improvement in energy resolution of soft x-ray RIXS has made it possible to study in detail both collective magnetic \cite{Ament09, Braicovich10, Betto17} and orbital \cite{Ulrich09, Benckiser15, Fabbris2016, Bisogni16} excitations in oxides. Additionally, RIXS is sensitive to excitations related to the electronic continuum, which are crucial for high-valence TMO, such as $R$NiO$_3$, where both local and itinerant excitations are possible \cite{Bisogni16, Hariki2018, Ament11}. Furthermore RIXS offers the typical advantages of resonant x-ray techniques, like bulk-sensitivity, element selectivity, and momentum-resolution, together with more detailed spectroscopic information than that provided by x-ray absorption spectroscopy (XAS) \cite{Ghiringhelli05}. In $R$NiO$_3$ several ordered phases coexist, making RIXS a powerful tool to probe these states simultaneously and in a site-selective manner \cite{Lu18}, and to gain access to several order parameters within the same experiment.\\
In this article we illustrate how RIXS can be used to simultaneously probe magnetic and bond order and the corresponding collective excitations in $R$NiO$_3$. In combination with theoretical models, we quantify the BO \footnote[1]{While we can differentiate between long- and short range BO, we cannot judge how long-range the order is using soft x-rays, as the corresponding BO Bragg reflection is not reachable.}
 as well as the magnetic exchange interactions at the basis of the unusual non-collinear spin spiral order of $R$NiO$_3$. We apply this methodology to thin films as well as to superlattices to explore the properties of $R$NiO$_3$ both in a bulk-like setting and in spatially confined layers with reduced dimensionality.

\section{Methods}\label{methods}
\subsection{Experimental Details}
For a systematic and quantitative study of the different ordering phenomena in $R$NiO$_3$, high-quality RIXS spectra with high-resolution are necessary. We therefore performed the RIXS experiments at the ID32 beamline of the European Synchrotron Radiation Facility using the ERIXS spectrometer \cite{Brookes19}. As a compromise between reasonable acquisition time and sufficient resolving power, the combined instrumental energy resolution was set to $\approx$ $50\,$meV full width at half maximum (FWHM). For the whole experiment we kept the incident photon polarization parallel to the scattering plane in order to enhance the magnetic response of the system. To measure the dispersive magnetic excitations, we varied the scattering angle in the range from $55\,^{\circ}$ to $135\,^{\circ}$, which corresponds to momentum transfer of 0.4 to 0.8$\,\text{\AA}^{-1}$ at the Ni $L_3$ edge at $853\,$eV. Additional resonant elastic x-ray scattering (REXS) experiments (Appendix\,\ref{azim}) were performed at the BESSY-II undulator beam line UE46-PGM1 at the Helmholtz-Zentrum Berlin.\\
High-quality thin films and superlattices (SLs) were grown by pulsed laser deposition. A $400\,\text{\AA}$ thick NdNiO$_3$ (NNO) film was grown on a $(001)_{\text{pc}}$-oriented SrTiO$_3$ substrate and has already been studied extensively as bulk representative \cite{Lu16, Lu18}. A LaNiO$_3$-LaAlO$_3$ (LNO-LAO) SL was grown on a $(001)_{\text{pc}}$-oriented LaSrAlO$_3$ (LSAO) and consists of 33 bilayers, each containing two pseudocubic unit cells (u.c.) of LNO and LAO. PrNiO$_3$-PrAlO$_3$ (PNO-PAO) SLs were grown on $(001)_{\text{pc}}$-oriented LSAO and [LaAlO$_3]_{0.3}$ $\times$ $[$Sr$_2$AlTaO$_6]_{0.7}$ (LSAT), with a ($2\,$u.c./$2\,$u.c.)$\times30$ and ($2\,$u.c./$4\,$u.c.)$\times20$ stacking of PNO and PAO, respectively. A NdNiO$_3$-NdGaO$_3$ (NNO-NGO) SL was grown on $(111)_{\text{pc}}$-oriented NdGaO$_3$ (NGO), which corresponds to the (101) direction in orthorhombic notation. The NNO-NGO SL comprises 4 bilayers with $3\,$u.c. of NNO ($16\,\text{\AA}$), separated by $2\,$u.c. of NGO ($12\,\text{\AA}$). In the NNO-NGO SL the unit cell is defined along the $(111)_{\text{pc}}$ direction. Details of all investigated samples can be found in Table\,\ref{properties}.

\begin{table*}[t]
\caption{Overview of sample properties. In the first column we refer to previous experiments on the same or nearly identical samples studied with other techniques. For the NNO-NGO SL, the unit cell is defined along the $(111)_{\text{pc}}$ direction and thus does not correspond to the unit cell used for the samples grown along $(001)_{\text{pc}}$.}
  \begin{ruledtabular}
\begin{tabular}{l c c c c c c c}
sample			& growth direction	  & stacking 	\hspace{4mm}	& $T_{\text{N}}/\text{K}$ 	 &  magnetic order & state at $T=20\,\text{K}$  \\
 \colrule
NNO thin film on STO \cite{Lu18, Lu16}		  &$(001)_{\text{pc}}$& - 			& 200 			&spiral		   & insulating \\

PNO-PAO SL on LSAO \cite{Hepting14, Wu15}		  &$(001)_{\text{pc}}$& $2\,$u.c./$2\,$u.c.	& 120 			&spiral		  & insulating \\
	
PNO-PAO SL on LSAT \cite{Hepting14, Wu15}		  &$(001)_{\text{pc}}$& $2\,$u.c./$4\,$u.c.	& 140 			&spiral		  & insulating \\

LNO-LAO SL on LSAO \cite{Boris11, Frano13}		  &$(001)_{\text{pc}}$& $2\,$u.c./$2\,$u.c	& 100			&spiral		  & metallic \\

NNO-NGO SL on NGO \cite{Hepting18}\hspace{8mm}&$(111)_{\text{pc}}$& $3\,$u.c./$2\,$u.c.	& 65 			&collinear		  & insulating \\

\end{tabular}
\end{ruledtabular}
\label{properties}
\end{table*}

 
\subsection{Double-Cluster Calculations}
To facilitate the quantitative analysis of our RIXS data, we calculate both XAS and RIXS spectra using the double-cluster model recently developed by Green \textit{et al.}\cite{Green16}. This model goes beyond the usual exact diagonalization approach based on a single Ni site surrounded by oxygen ligands \cite{Haverkort12}. Instead, the double-cluster formalism comprises two NiO$_6$ clusters to explicitly include LB and SB sites, thus reproducing the elementary building block of the rocksalt pattern of alternating octahedra in $R$NiO$_3$. Each cluster is described by a standard multiplet ligand field Hamiltonian including the full Coulomb interactions and the necessary orbital degeneracies \cite{Ballhausen, Fuersich18}. The two clusters are then coupled by hybridization operators with $O_h$ symmetry. 
The calculations are performed with the exact diagonalization code Quanty\cite{Haverkort12, Haverkort14, Lu14}.\\
The double-cluster model considers several key features of the valence electron system of $R$NiO$_3$, including negative charge transfer energy, Coulomb interactions, orbital degeneracies and bond disproportionation. Importantly, the latter cannot be incorporated in the commonly used single-cluster models. The negative charge transfer picture is essential to reproduce the \textit{self-doped} ground state. In general, the charge transfer energy $\Delta$ describes the cost to transfer one electron from the ligand to the TMO $3d$ band \cite{Zaanen85, Mizokawa91}. In $R$NiO$_3$ $\Delta$ is negative, therefore one hole is doped into the ligand states leading effectively to an O $2p$ - O $2p$ gap \cite{Mizokawa00, Bisogni16}. We follow the conventions of Refs. \onlinecite{Green16, Bisogni16} and define $\Delta$ as the energy difference between the top of the ligand band and the bottom of the $3d$ band. Consequently, $\Delta$ gives the energy separation of $d^7\underline{L}^0$ and $d^8\underline{L}^1$ configurations in the case of  $R$NiO$_3$. In order to obtain the exact multiplet structure and its spectroscopic fingerprint, it is of crucial importance to include Coulomb interactions as well as orbital degeneracies, as discussed in the literature \cite{deGroot, Cowan}.\\
The interaction between clusters is quantified by the inter-cluster mixing $V_{\text{I}}$, which is proportional to the ratio between intra- and inter-cluster hopping, that are, respectively, the hopping within a single NiO$_6$ cluster and between the two clusters. In this way charge fluctuations among two neighboring NiO$_6$ octahedra are explicitly incorporated in the formalism. One can thus achieve several configurations beyond those of the classical single-cluster picture, thereby accounting for the highly covalent character of the ground state of  $R$NiO$_3$. Most importantly, the calculations show that the high-temperature ground state is dominated by the self-doped $d^8\underline{L}^1$ configuration.\\ 
In the low-temperature insulating ground state a bond disproportionation $\delta d$ is introduced. This parameter is defined as the displacement of the oxygen position along the Ni-O-Ni bonds from the mean value without bond disproportionation, following the definitions in  Ref.\,\onlinecite{Green16, Lu18}, where the double-cluster approach was employed. For comparison with previous studies, we note that the bond disproportionation can also be quantified as the difference between short and long Ni-O bond lengths \cite{Lu16, Medarde97}, which doubles the value of $\delta d$ compared to our definition. We incorporate the breathing distortion to the model by adjusting mixing and crystal-field terms according to Harrison's rules \cite{Harrison83, Johnston14}. Within the double-cluster model the alternating octahedra with $d^8\underline{L}^0$ $(S\approx1)$ and $d^8\underline{L}^2$ $(S\approx0)$ configuration follow naturally.\\

\begin{figure}
\includegraphics[width=0.9\columnwidth]{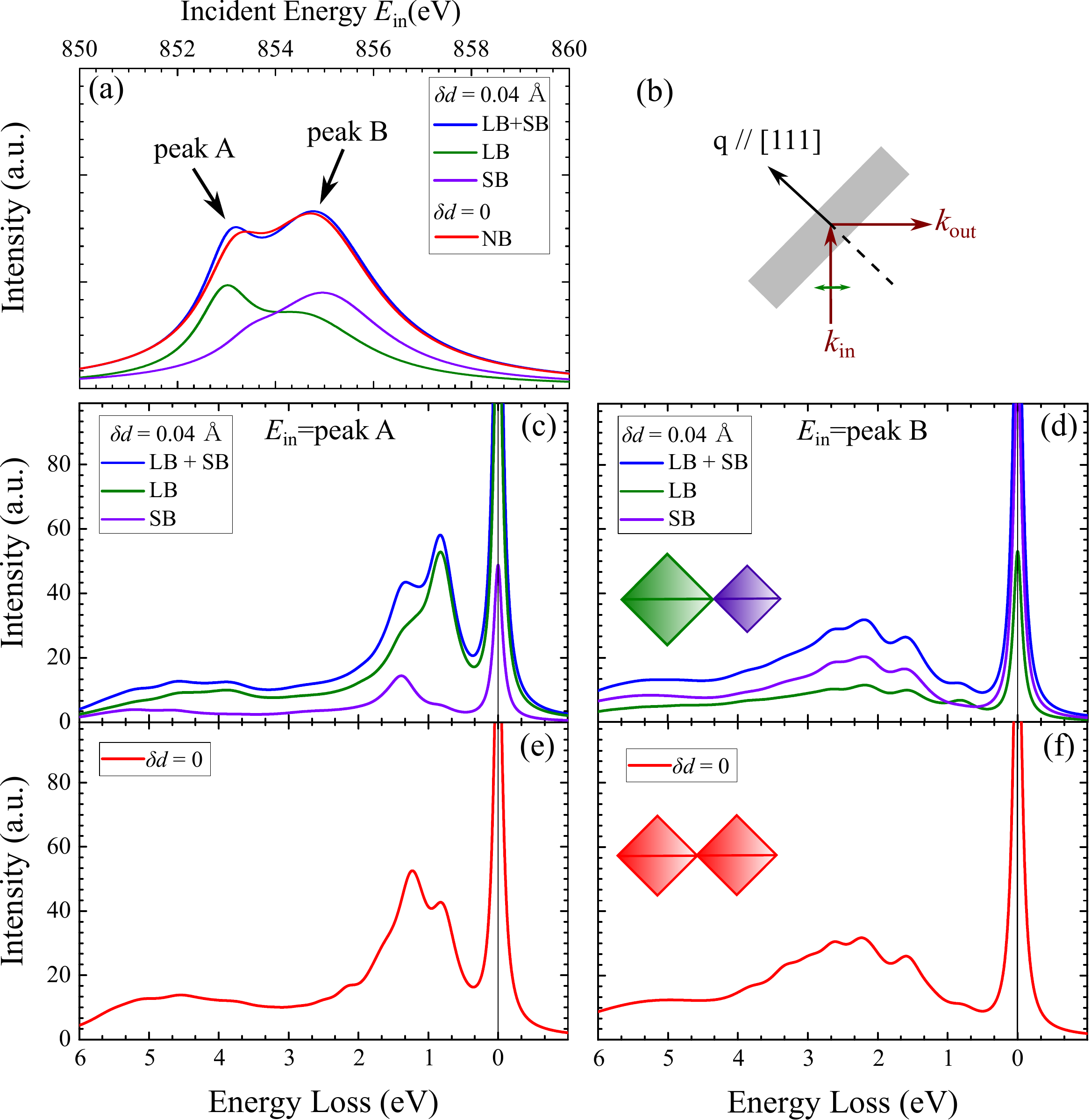}%
\caption{Overview of calculated XAS and RIXS spectra within the double-cluster model. Panel (a) gives the theoretical XAS at the Ni $L_3$ edge calculated with (blue) and without (red) bond-disproportionation, and defines the incident energies: peak A and B. (b) Scattering geometry for the RIXS experiment. (c) and (e): RIXS spectrum with incident photon energy tuned to peak A, as defined in panel (a). (d) and (f): RIXS spectrum with incident energy tuned to peak B. Spectra in panels c,d (e,f) were calculated with (without) bond disproportionation. The spectrum for the low-temperature bond-ordered state is the sum of the spectra corresponding to LB (green) and SB (purple) octahedra. The sketches in panels (d) and (f) illustrate the BO below and above the MIT transition. For clarity, the octahedral rotations are omitted.}
\label{Quanty}
\end{figure}

In the next sections we use the double-cluster model to calculate XAS and RIXS  spectra. Hereafter we introduce the procedure we followed and the output of the calculations by showing a typical example for the energy ranges and scattering geometries used in this work [Fig.\,\ref{Quanty}(a) and (b)]. For this survey, we use the parameters given in  Ref.\,\onlinecite{Lu18}, namely $V_{\text{I}}=0.33$ and $\delta d=0.04\,\text{\AA}$ at a momentum transfer of  $q_{\text{[111]}}=0.211$ [short for $q_{\text{[111]}} = 0.211(1, 1, 1)$]. The calculated spectra are broadened to account for experimental resolution and life-time effects. As a first step, we inspect the XAS spectrum to choose the resonant incident energies at which the following RIXS experiment/calculation is carried out.  Fig.\,\ref{Quanty}(a) illustrates the Ni $L_3$ XAS spectra calculated within the double-cluster model. In the non-disproportionated case, we find a two-peak structure at the Ni $L_3$ edge due to dynamic charge order, \textit{i.e.} charge fluctuations between the clusters. In the disproportionated state the spectrum consists of contributions from LB and SB octahedra. In the presence of BO, the two peaks change only weakly, but can now be attributed to strictly different static contributions, with peak A arising predominantly from the LB site and peak B arising almost equally from LB and SB sites \cite{Green16}. The double-peak structure as well as the small energy shift between zero and nonzero bond-disproportionation found in the calculation is a distinct property of $R$NiO$_3$ \cite{Piamonteze05, Freeland16}. Already when considering the XAS spectra, it is evident that the double-cluster model reproduces the experiment [Fig.\,\ref{PNO}(a)] much better than the conventional single cluster model, which displays only one sharp peak at the Ni $L_3$ edge \cite{Wu13, Benckiser11}. Consequently, to obtain an accurate description of the RIXS spectra, which exhibit far more fine details and features, it is essential to adopt the double-cluster model. \\
We now turn to the discussion of the  RIXS calculation.  Fig.\,\ref{Quanty} gives an overview of the calculated RIXS spectra within the double-cluster model for both $\delta d\neq0$ and $\delta d=0$ (corresponding to low- and high-temperature phases, respectively) as well as for incident energy tuned to peak A and peak B, as defined in the XAS spectra in panel (a). Irrespective of $\delta d$, the shape of the spectra measured with incident energy tuned to peak A is quite different from the ones at peak B. While the spectrum at peak A displays sharp features around $1\,\text{eV}$, we identify a broad component around $2\,\text{eV}$ at peak B. This observation implies a different origin of the excitations at peak A and B, thereby suggesting coexistence of bound and continuum excitations within one material. Indeed, Bisogni \textit{et al.} \cite{Bisogni16} attributed the different features to local and band-like excitations by carefully monitoring their energy and temperature dependence. At peak A mostly bound $dd$ excitations are observed, whereas at peak B the RIXS spectrum has a predominant contribution from band-like fluorescence decay. In addition, for both incident energies, charge-transfer excitations lead to a broad high-energy background around $4\,\text{eV}$.\\
Firstly, we analyze the spectra calculated with nonzero bond-disproportionation describing the insulating low-temperature state. In the bond-ordered phase, the calculated spectrum comes from the sum of the contributions arising from the  LB and SB site. To gain a deeper understanding of the energy dependent excitations, we disentangled the contributions from LB and SB octahedra by separately plotting their individual spectra [panels (c) and (d) in  Fig.\,\ref{Quanty}]. Interestingly, the spectrum measured at peak A consists mostly of contributions from the LB site, corresponding to the expanded octahedron, while the SB site only adds minor spectral weight. Similar conclusions have been reached in  Ref.\,\onlinecite{Ruppen2015}. The distribution of LB and SB contributions changes substantially when tuning the incident energy to peak B. We find almost equal contributions from LB and SB octahedra, in close analogy to the XAS spectrum at low temperatures.\\
Secondly, we take a closer look at the temperature dependence of the calculated spectra, exemplified in the comparison between $\delta d=0$ [high-temperature phase, panels (e),(f)] and $\delta d\neq0$ [low-temperature phase, panels (c),(d)]. At peak A, we observe changes in the $dd$ excitations around $1\,\text{eV}$. The spectral weight of the double peak structure shifts towards the high-energy side. In stark contrast, at peak B we recognize no obvious difference between zero and nonzero BO.\\
From this first overview of the double-cluster calculation, we conclude that clear signatures of BO can be gleaned from RIXS spectra measured with incident energy tuned to peak A. In addition, as known from our previous work \cite{Lu18}, dispersive spin excitations can only be measured at peak A. Therefore, we will focus on spectra measured at peak A in the following sections.

\section{Quantifying the bond-disproportionation in \emph{R}N\lowercase{i}O$_3$} \label{BO}
\subsection{Bond order in $R$NiO$_3$ films and superlattices with $R\neq\text{La}$} \label{bond order in PrNiO$_3$-based superlattices}

In order to get a systematic overview, we measured a film and three SLs with different periodicities and rare-earth species, specifically Nd and Pr. This allowed us to study the BO as a function of tolerance factor, \textit{i.e.} rare-earth ion radius, as well as in the limit of two-dimensional confinement. The rare-earth ion radius has a strong influence on the onset temperature and strength of the BO, as it effectively controls the bandwidth via the Ni-O bond angle and the consequent Ni $3d$ and O $2p$ hybridization \cite{Torrance92, Alonso00}. To access the BO parameter, RIXS is an excellent tool as it is sensitive to dipole forbidden inter-orbital $dd$ excitations, which are strongly influenced by the electronic reconstruction associated with the bond-ordered phase.

\begin{figure*}
\includegraphics[width=0.99\textwidth]{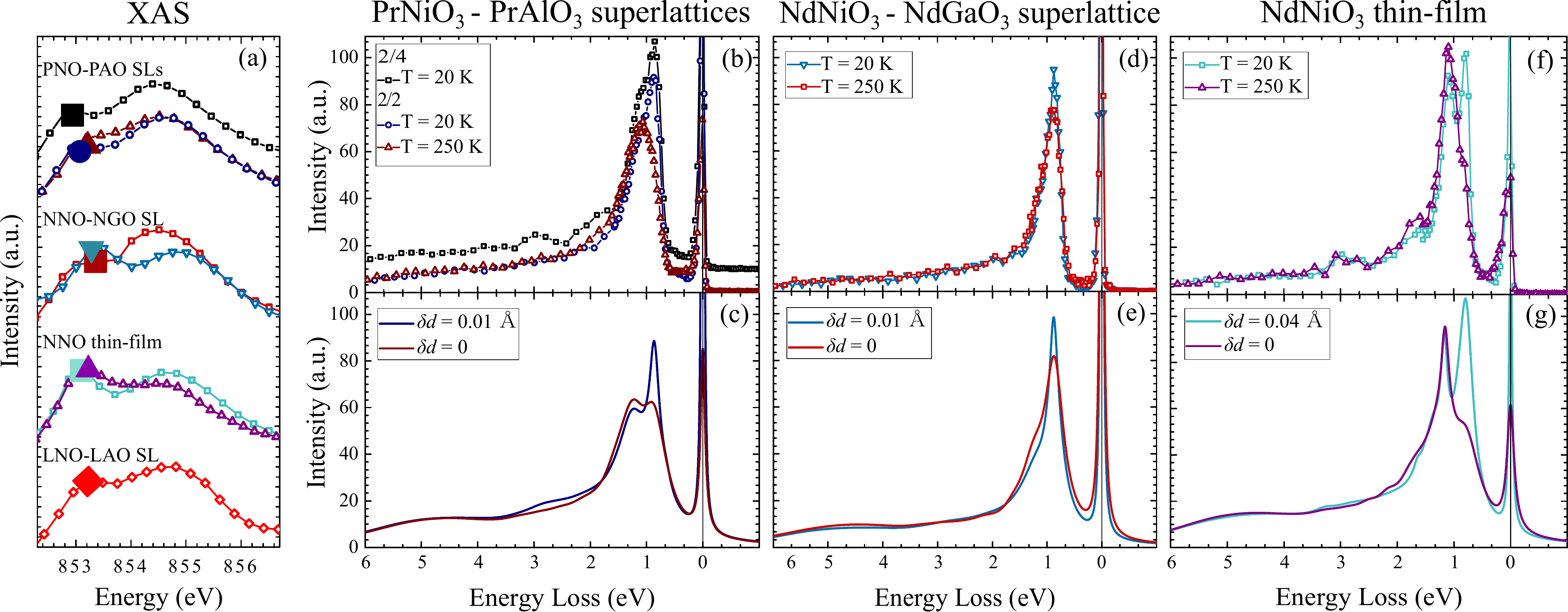}
\caption{(a) XAS spectra for all investigated samples. The big symbol indicates the incident energy for the RIXS experiment, \textit{i.e.} the peak A. For the LNO-LAO SL we subtracted the La $M_4$ line. (b) RIXS spectra for two PrNiO$_3$-based superlattices with different stacking periodicities. The SL with 2/4 stacking is offset for clarity. (c) Double-cluster calculation for the PNO-PAO SLs.  (d) RIXS spectra for the NdNiO$_3$-NdGaO$_3$ SL with the corresponding double-cluster calculation in (e). For both SLs we find a bond-disproportionation of $\delta d=0.01\,\text{\AA}$. (f) Experimental and (g) calculated spectra for a NdNiO$_3$ reference thin film. Panel (f) and (g) are reproduced from  Ref.\,\onlinecite{Lu18}.}
\label{PNO}
\end{figure*}

Fig.\,\ref{PNO}(b) shows spectra for two PNO-PAO SLs measured with an incident energy tuned to peak A at $q_{\text{[111]}}=0.211$ for $T=20\,$K and $T=250\,$K. For both PNO-PAO SLs with different stacking periodicity we observe almost identical spectra indicating a similar electronic structure. Comparing these data to the spectra from a NNO SL and a film [Fig.\,\ref{PNO} (d) and (f), respectivly] measured in the same conditions, we observe remarkable differences for both the bond-ordered and the non bond-ordered phase. While most of the spectral weight for all systems is centered around $1\,\text {eV}$, the features in the NNO film are much sharper compared to the SLs. Additionally, the double peak structure in the NNO film is more pronounced.\\
To understand these observations in a quantitative manner, we performed double-cluster calculations as described in Section \ref{methods}. We used the parameters given in Ref.\,\onlinecite{Lu18} and solely optimized the bond-disproportionation $\delta d$ and inter-cluster mixing $V_{\text{I}}$ to reproduce the experimental data. The calculated spectra that best describe the experimental data are shown in  Fig.\,\ref{PNO}, panels (c), (e) and (g).\\
We first discuss the NNO film as reference for the bulk phase. There we find $V_{\text{I}}=0.33$ and $\delta d=0.04\,\text{\AA}$ \cite{Lu18}. The bond disproportionation parameter is in excellent agreement with previous values from x-ray scattering at the Ni $K$ edge \cite{Lu16, Staub02} and powder diffraction measurements \cite{Garica-Munoz09}. The consistency with the literature validates our approach and shows that our method is applicable to $R$NiO$_3$. We emphasize that the experimental RIXS spectra are much better reproduced with the double-cluster model than with the standard single cluster model, used in  Ref.\,\onlinecite{Bisogni16} as it allows a quantitative determination of the BO parameter. \\
For the PNO-PAO superlattices we find the same value for the inter-cluster mixing  $V_{\text{I}}=0.33$ and a lower bond-disproportionation $\delta d$=$0.01\,\text{\AA}$ in the low-temperature phase. In general, the reduced BO in bulk PNO can be explained as a consequence of the greater overlap between Ni $3d$ and O $2p$ orbitals, in comparison to NNO, which also reduces the critical temperature for the MIT. Powder diffraction measurements revealed a bond-disproportionation of $\delta d$=$0.026\,\text{\AA}$ for bulk PNO \cite{Medarde08}. The even lower value for $\delta d$ in PNO-based superlattices found here can be attributed to two-dimensional confinement and pinning of the oxygen positions in the nickelate layers at the interfaces with the buffer layer, which further increases the bandwidth \cite{Wu13, Boris11}. Raman scattering showed that the bond order in PNO-PAO SLs can be completely suppressed in compressively strained SLs and a pure metallic spin-density wave was found \cite{Lee11, Hepting14}. However, even though the 2/2 PNO-PAO SL investigated in this study is under compressive strain, we do not observe a complete suppression of bond order. Since RIXS is a more sensitive probe of BO than the detection of extra phonon modes via Raman scattering,  it would have been challenging to conclusively identify the weak BO distortion reported here in the Raman experiments. 
On the other hand, the BO parameter is possibly not completely suppressed due to partial relaxation of the rather thick SL used in the present work (see Appendix\,\ref{XRD}) and in  Ref.\,\onlinecite{Frano13AdvMat} compared to the one in  Ref.\,\onlinecite{Hepting14}.\\
The BO amplitude in Nd-based SLs can be greatly reduced due to spatial confinement, similar to the example of the Pr-based SLs. As an example we refer to a NNO-NGO SL grown on a (111)$_{\text{pc}}$ oriented substrate shown in  Fig.\,\ref{PNO}, panels (d) and (e), where we find a reduced bond-disproportionation of $\delta d$=$0.01\,\text{\AA}$. The different shapes of the $dd$ excitations in the NNO-NGO SL and the NNO film for $\delta d =0$ is related to the different cross sections for the (111)$_{\text{pc}}$ and (001)$_{\text{pc}}$ oriented samples \cite{Moretti11}.\\
Although thin-films and SLs give rise to completely different shapes of the RIXS spectra, the double-cluster approach with tuned parameters describes both systems very well (lower panels of  Fig.\,\ref{PNO}). This gives evidence that the double-cluster model, and specifically the negative charge transfer scenario, is an excellent picture to describe the local physics in $R$NiO$_3$ and suggests that holes in the oxygen ligands  are a key ingredient to understand the MIT. Our approach can be readily applied to all other $R$NiO$_3$ with $R\neq\text{La}$, where BO essentially governs the insulating phase. However, it is interesting to ask whether the double-cluster model can be adopted also for the correlated metal LaNiO$_3$, the only compound in the $R$NiO$_3$ family which shows BO neither in bulk nor in heterostructures.

\subsection{Breathing-type fluctuations in LaNiO$_3$}

\begin{figure}
\includegraphics[width=0.5\columnwidth]{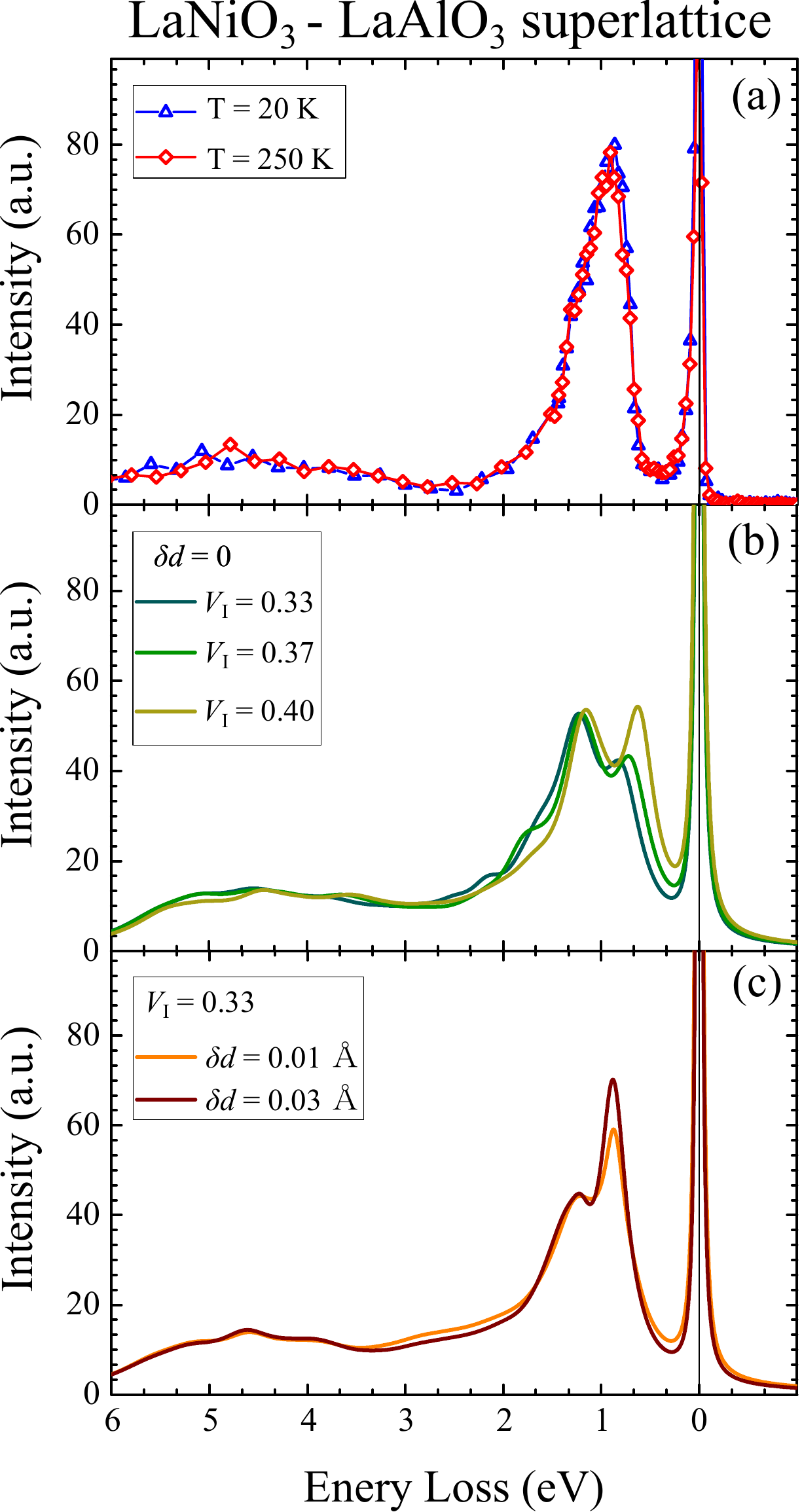}%
\caption{(a) RIXS spectra for a LNO-LAO SL. (b) Calculation without bond-disproportionation $\delta d$ and adjusted inter-cluster mixing $V_{\text{I}}$. (c) Calculation with small bond-disproportionation. The spectra calculated with finite $\delta d$ reproduce the experiment much better. $V_{\text{I}}$ is kept the same as in the PNO and NNO-bases samples.}
\label{LNO}
\end{figure}

LaNiO$_3$ (LNO) can be considered as an exception among the $R$NiO$_3$ compounds as it stays paramagnetic and metallic at all temperatures. This is due to the fact that $R$=La is the largest rare-earth ion in the $R$NiO$_3$ family, so that in LNO the Ni-O bonds are rather straight, resulting in an increased hybridization of Ni $3d$ and O $2p$ orbitals. However, it has been demonstrated that by confining the active LNO layers towards a planar two-dimensional limit, one can induce AFM order \cite{Boris11}. This can be viewed as a spin-density-wave (SDW) ground state in the absence of BO\cite{Lee11, Lu17}. 
\begin{figure*} 
\includegraphics[width=1\textwidth]{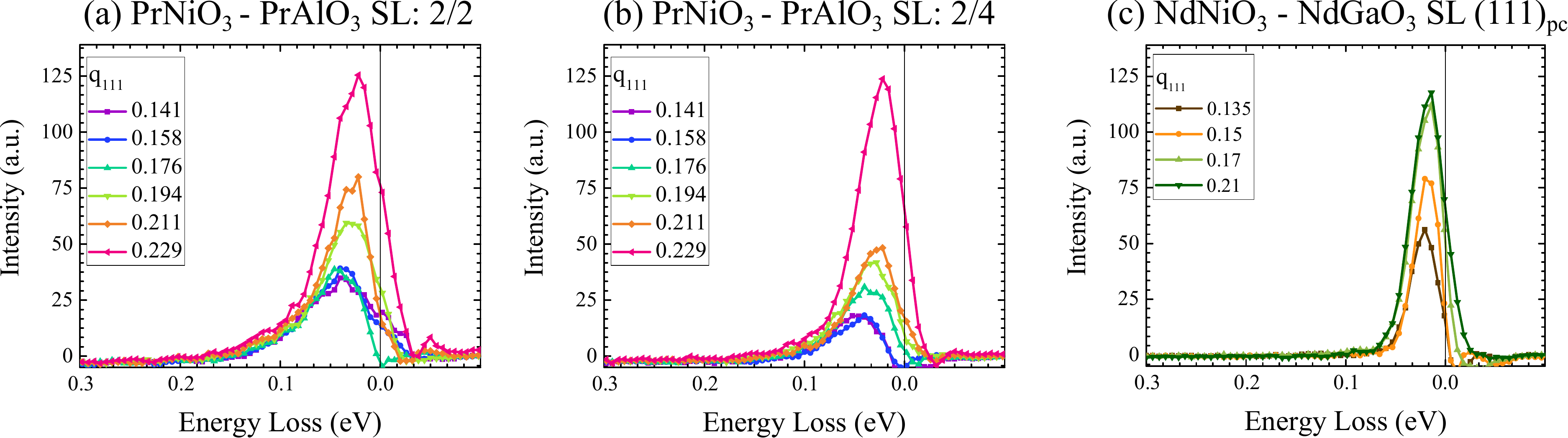}%
\caption{Dispersive magnetic excitations in PNO-based heterostructures after removal of elastic contribution and subtraction of high-temperature data, along the lines of  Ref.\,\onlinecite{Lu18}. (a) PNO-PAO ($2\,$u.c./$2\,$u.c) SL on LSAO. (b)  PNO-PAO ($2\,$u.c./$4\,$u.c) SL on LSAT. (c)   NNO-NGO ($3\,$u.c./$2\,$u.c) SL on NGO $(111)_{\text{pc}}$.}
\label{Magnon}
\end{figure*}
To gain a deeper understanding of the electronic mechanism inducing the SDW ground state and to test the double-cluster model against a highly correlated metal, we employ the same approach described in the previous section. The RIXS spectra are measured at peak A with the scattering vector $q_{\text{[111]}}=0.211$ and for two temperatures, $T=20\,$K and $T=250\,$K. As illustrated in  Fig.\,\ref{LNO}, the spectrum of the LNO-LAO SL does not evolve with temperature above $1\,$eV energy loss. Therefore, we can safely exclude long-range static bond order in accordance with our previous resonant diffraction experiments, where  no BO Bragg reflections could be observed at the Ni $K$ edge \cite{Lu16}.\\
Remarkably, the spectra from the LNO-LAO SL are quite similar to those from PNO-PAO SLs at low temperature which could be modeled with small but non-vanishing bond-disproportionation. We therefore tried to reproduce the experimental findings by using the same double-cluster model as in Sec. \ref{bond order in PrNiO$_3$-based superlattices}. We account for the larger bandwidth of LNO by increasing the inter-cluster mixing $V_{\text{I}}$, while maintaining the constraint of absent bond order ($\delta d=0$). The calculation fails to reproduce the experimental RIXS lineshape even on a qualitative level [see  Fig.\,\ref{LNO} (b)]. As a next step, we allowed for small local breathing-distortions $\delta d\neq0$. The theoretical spectra calculated with nonzero $\delta d$ reproduce the experiment much better. While some discrepancies between the numerical and experimental data remain, the results indicate short-range order in the form of transient BO-like distortions, \textit{i.e.} breathing-type fluctuations of the NiO$_6$ octahedra. The BO fluctuations are observable by RIXS, because they are much slower than the RIXS scattering process itself. A complete quantitative description of the RIXS spectrum in this situation remains a challenge for future theoretical work.\\
Several independent findings point out the importance of BO fluctuations in the description of  $R$NiO$_3$, and in particular for $R=\text{La}$ \cite{Medarde09, Johnston14, Lau13,  Lu17}. Recent experiments based on the pair distribution function (PDF) method have found evidence for two nonequivalent Ni sites in LNO even in the metallic phase \cite{Shamblin18, Li16}. Some studies even suggest that bond-length fluctuations are present in all  $R$NiO$_3$ at high temperature, thereby classifying the metallic state in $R$NiO$_3$ as a polaronic liquid. The MIT and the associated BO can then be explained in terms of stabilization/freezing of the pre-formed fluctuating rock-salt pattern of octahedra from the metallic state \cite{Shamblin18}. Additional evidence for charge/bond fluctuations can be found in the Fermi surface superstructure with wavevector $\textbf{Q}_{\text{BO}}=(\nicefrac{1}{2}, \nicefrac{1}{2},\nicefrac{1}{2})_{\text{pc}}$ observed by angle resolved photo emission in metallic LNO \cite{Yoo15}. Together with our RIXS data, these results suggest that BO fluctuations are essential for the theoretical description of LNO.

Using high resolution RIXS in combination with a double-cluster model, we discriminated between long- and short-range bond order and quantified the bond-disproportionation in several representatives of $R$NiO$_3$. In the following we elucidate the effect of different BO strengths on the spin excitations in $R$NiO$_3$. 
 
\section{Spin excitations in \emph{R}N\lowercase{i}O$_3$}

\begin{figure}
\includegraphics[width=0.42\textwidth]{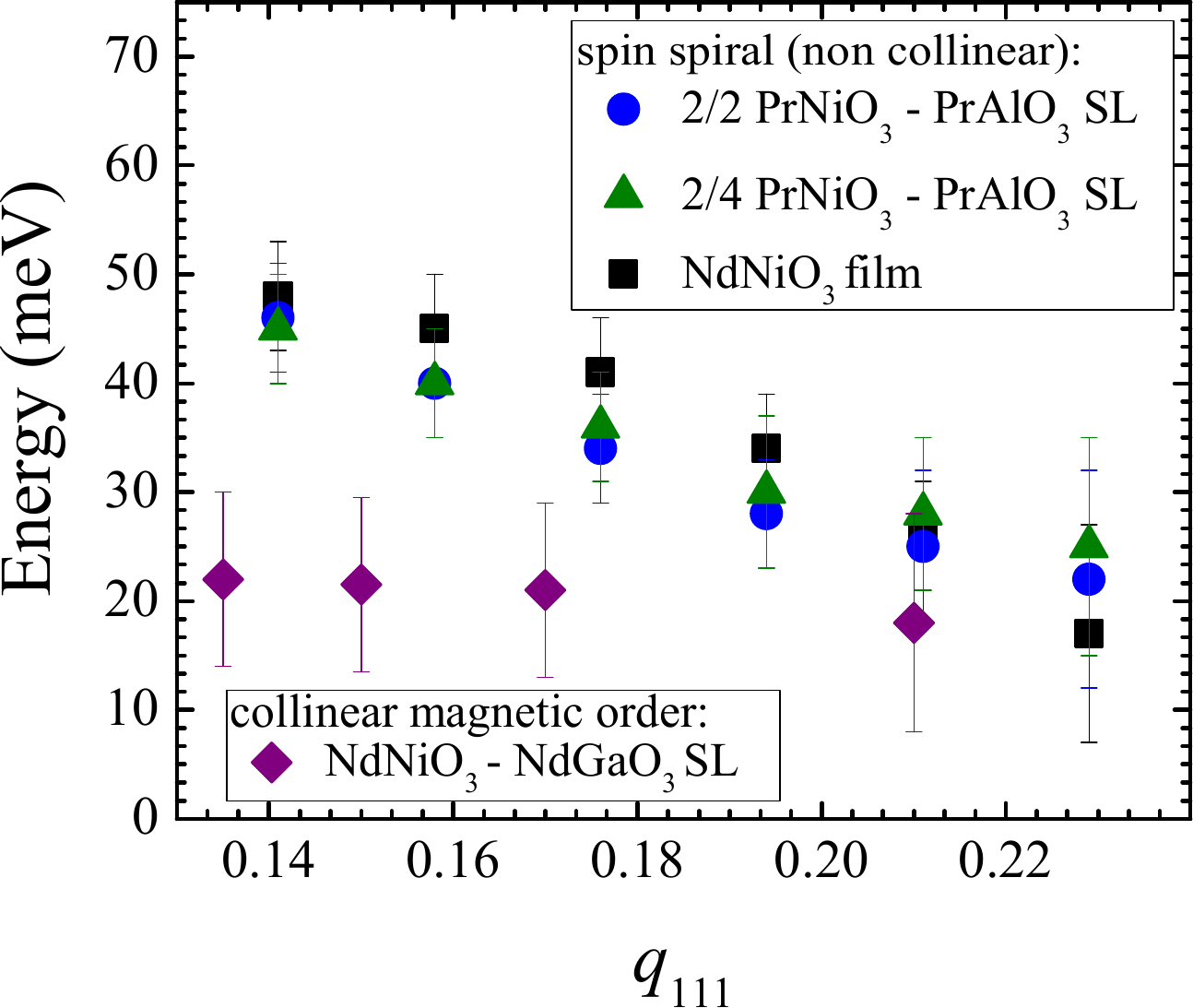}%
\caption{Magnon dispersion of PNO-based superlattices compared to the one from the NNO-NGO $(111)_{\text{pc}}$ superlattice (extracted from Figure \ref{Magnon}). In addition we show the magnon dispersion for a NNO thin film which is taken from  Ref.\,\onlinecite{Lu18}.}
\label{dispersion}
\end{figure}

As far as collective spin excitations are concerned, one can consider three different cases of magnetic order in $R$NiO$_3$ compounds. For insulating bulk-like films the robust BO is a prerequisite for the spin spiral that appears at $T_{\text{MIT}}$ or lower temperature depending on the rare-earth $R$ \cite{Scagnoli08}. In spatially confined systems, as realized in SLs grown along the $(001)_{\text{pc}}$ direction, the spiral magnetic order can develop with weak or absent BO \cite{Hepting14, Wu15, Frano13}. Among this SL family the LNO-LAO is special, as the system remains metallic while developing the spiral order \cite{Lu16, Frano13}. Additionally, collinear magnetic order can be found in SLs with $(111)_{\text{pc}}$ orientation \cite{Hepting18} matching the magnetic propagation vector. Previously we have studied the magnon excitations in insulating bulk-like films \cite{Lu18}. Here we will focus on the latter two cases, namely the $(001)_{\text{pc}}$ SLs with spiral magnetic order and reduced BO, and the $(111)_{\text{pc}}$ SL with collinear magnetic order.\\
For the investigation of magnetism in  $R$NiO$_3$, we focus on the low-energy part of the spectra already shown in the previous section. We use the same examples from Sec. \ref{BO} for strong and weak BO, namely the NNO film and the PNO-PAO SLs with two different stacking periodicities. For $R$NiO$_3$ with $R$=Nd, Pr bond and magnetic order set in at the same temperature ($T_{\text{MIT}}=T_{\text{AFM}}$).\\
We use our recently developed approach to measure the dispersive spin excitations in several PNO- and NNO-based heterostructures \cite{Lu18}. To single out the purely magnetic signal, contributions from elastic scattering and other low-energy excitations were subtracted from the spectra, following the procedure presented in  Ref.\,\onlinecite{Lu18}. The elastic line is given by a Gaussian peak with $\text{FWHM}=50\,\text{meV}$ (experimental resolution) at zero energy loss. The spectrum measured above the magnetic ordering temperature gives the non-magnetic low-energy excitations, dominated by phonons.\\
After subtracting the high-temperature inelastic spectra from the low-temperature inelastic data one is left with well-defined dispersive magnetic features. Magnon dispersions for different stacking periodicities, rare earth ion and substrate orientation can be seen in  Fig.\,\ref{Magnon}. For all samples, we observe an increase in the magnetic spectral weight as we move towards \textbf{Q}$_{\text{AFM}}$. 
Moreover, the energy of the spin excitations disperses from approximately 50$\,$meV to 20$\,$meV as the scattering vector gets closer to \textbf{Q}$_{\text{AFM}}$ for the PNO-based SLs, while the magnon bandwidth is clearly reduced for the NNO-NGO SL. This is further illustrated by the extracted magnon dispersion shown in Fig.\,\ref{dispersion}.\\
The variations in the magnon dispersion can be related to the microscopic spin structure. The PNO-PAO SLs host the well known non-collinear AFM spin spiral, extensively studied for example by Frano \textit{et al.}\cite{Frano13}. In contrast, the NNO-NGO SL orders in the recently discovered collinear pattern (see appendix \ref{azim})\cite{Hepting18}. \\
We first focus on the PNO-PAO SLs, which show a spiral ground state and a similar dispersion as the NNO film from   Ref.\,\onlinecite{Lu18}. Both the ground state and the low-energy excitation spectrum of NNO were explained by a $J_1-J_2- J_4$ model with exchange interactions between nearest-, second-nearest, and fourth-nearest-neighbor Ni spins. $J_1$ is anomalously small due a strong competition between the AFM super-exchange and the ferromagnetic (FM) double-exchange interactions. The AFM ordering within one sublattice of equally sized octahedra and magnetic moments follows from the $J_2$ coupling, which is dominated by superexchange interactions. Low-energy charge fluctuations between nearest-neighbor sites lead to a FM double-exchange interaction, which is optimized for an angle of 90$^\circ$ between adjacent spins. The close similarity between the magnon dispersions of the NNO film and of the PNO-PAO SLs shows that the magnetism in the two-dimensional limit can be explained by the model developed for bulk NNO with similar exchange coupling constants (Fig.\,\ref{dispersion}). According to this model, the strongest exchange interactions $J_2$ and $J_4$ connect spins within the same sublattice of the BO state, whereas the nearest-neighbor interaction is weaker and does not substantially affect the measured magnon dispersion \cite{Lu18}. The model therefore naturally explains the observed insensitivity of the spin dynamics to the BO parameter. We therefore conclude that our magnetic model can be applied to a wide range of  $R$NiO$_3$ thin films and heterostructures with different electronic and structural properties. In particular, the increased metallicity (and bandwidth) obtained from a combination of compressive strain and a larger rare-earth ion does not have a significant impact on the magnon dispersion. The spin spiral in $R$NiO$_3$ is thus essentially unperturbed by a modulation of the BO.\\

\begin{figure}
\includegraphics[width=0.42\textwidth]{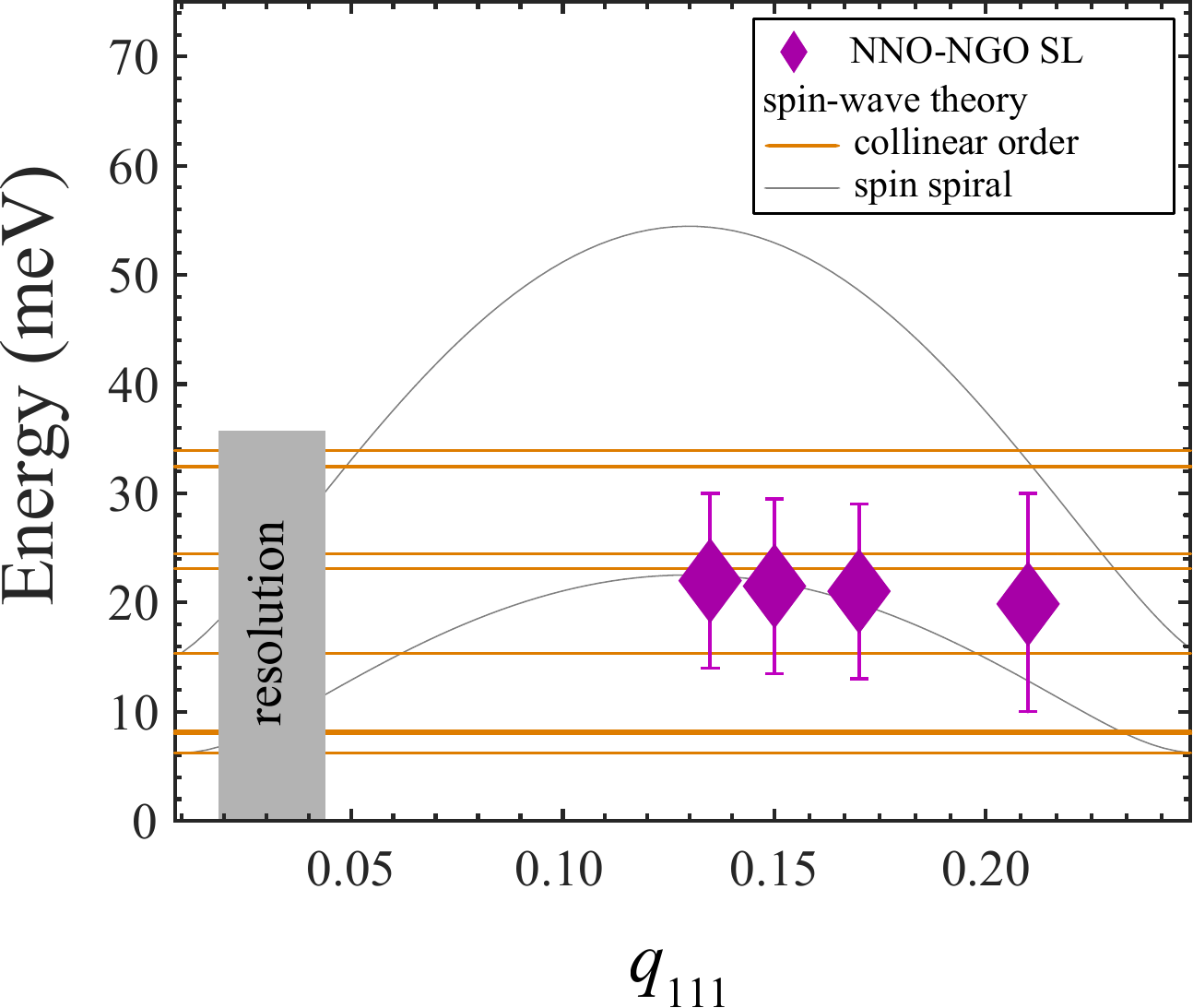}%
\caption{Linear spin-wave theory for a collinear magnetic structure. The magnetic supercell comprises 8 ML NNO separated by non-magnetic NGO. Orange lines show the low-energy eingemodes of the system. The purple diamonds represent the measured data, while the gray box indicates the resolution limit in the RIXS experiment. Gray lines indicate the calculated dispersion of magnons in the spiral state of a bulk-like NNO film for reference \cite{Lu18}.}
\label{SpinW}
\end{figure}

We now proceed to the collinearly ordered NNO-NGO SL, where the magnon energy near the magnetic zone boundary is reduced by a factor of 2 (see  Fig.\,\ref{dispersion}). According to  Ref.\,\onlinecite{Hepting18} the different spin structure is a consequence of truncated exchange bonds along the magnetic ordering vector inherent to the particular SL geometry. We therefore construct a magnetic supercell comprising 3 u.c. NNO, separated by the non-magnetic NGO, and stacked along the $[111]_{\text{pc}}$ direction. We start from the magnetic structure determined by the REXS experiments (see appendix \ref{azim}) as well as the bulk exchange parameters and set the bond-disproportionation to $\delta d=0.01\,\text{\AA}$, \textit{ i.e.} $S_{\text{LB}}=0.55$ and $S_{\text{SB}}=0.45$ as suggested by the double cluster calculation (see  Fig.\,\ref{PNO}). The ground state and the magnetic dispersion are numerically computed using the SpinW software package \cite{Toth15}. The result is shown in  Fig.\,\ref{SpinW}, where the low-energy eigenmodes are indicated by orange lines. These modes are dispersionless in the $(111)_{\text{pc}}$ direction and their energies are lower than the zone-boundary energy of the bulk dispersion (gray lines). In order to compare the calculated modes to the experimental data, one has to consider the experimental resolution indicated by the gray bar. It is evident that within the current resolution the predicted splitting of the low-energy modes cannot be resolved. However, the overall energy scale and the lack of dispersion in the $(111)_{\text{pc}}$ direction are in good agreement with the model calculation.\\

In the LNO-LAO SL, where magnetic order was previously observed by muon spin rotation and resonant elastic x-ray scattering \cite{Boris11,  Frano13}, we detected an increase in spectral weight upon approaching \textbf{Q}$_{\text{AFM}}$ both in the elastic and in the inelastic channel, but no dispersive feature. This may be a consequence of heavy damping of the magnon modes by incoherent particle-hole excitations in the metallic sample. We note, however, that these experiments were hampered by strong self-absorption due to the proximate La $M$ edge, so that no firm conclusion on the absence of pronounced magnon modes in the RIXS spectra could be reached.

\section{Conclusions and Outlook}
In summary, we used high-resolution RIXS to simultaneously probe the bond and magnetic order in a representative selection of $R$NiO$_3$ thin films and superlattices. Firstly, we showed that RIXS in conjunction with multiplet calculations in the framework of a double-cluster model can serve as a highly sensitive probe of BO. We found a variety of long-range BO strengths for bulk-like films and for SLs of $R$NiO$_3$ with different rare-earth ions $R$. Additionally, we observed indications of fluctuating short-range BO in LNO-LAO SLs. Secondly, we investigated the magnetic properties of the same samples and established that the spin spiral magnetism is a robust order, which develops in most $R$NiO$_3$ systems irrespective of the BO strength. We also showed that the model recently developed in  Ref.\,\onlinecite{Lu18} for the magnetic excitations in bulk $R$NiO$_3$ provides an accurate description of the magnon dispersions in SLs with non-collinear magnetic order. In the case of SLs with collinear magnetic order, on the other hand, we find an essentially flat dispersion with reduced magnon energies. This observation is explained by a spin-wave theory with the same interaction parameters adapted for the particular SL geometry. Our approach determines bond and magnetic order on a quantitative level, which is of great importance to understand the feedback between these two different ordering phenomena. According to recent theoretical work, this interplay is a key factor for the emergence of exotic phases like multiferroicity \cite{vandenBrink08, Giovannetti09} or even potentially superconductivity \cite{Chaloupka08, Hansmann09} in $R$NiO$_3$.\\
In the future, it would be of great interest to study systems with a single active magnetic nickelate layer, in which the conventional non-collinear spin spiral cannot develop. The spin excitations measured by RIXS could give valuable insight to develop models for genuinely two-dimensional magnetism in $R$NiO$_3$. 

Our approach to analyze the intra-orbital excitations with a double-cluster model could be used as a reference for studies of other high-valence and highly covalent TMOs \cite{Balandeh17, Khazraie18}. The accurate determination of bond order parameters by high-resolution RIXS is especially relevant for other materials with BO such as manganites \cite{Efremov2004} and tellurides \cite{ Takubo2014}, which can  be characterized in a similar way.

\begin{acknowledgments}
We thank G. Khaliullin, R. Green and G. Sawatzky for fruitful discussions. We acknowledge financial support from the German Science Foundation under Grant No. TRR80. N. B. B. would like to thank Diamond Light Source, Didcot, UK, for hosting him during part of 2019. 
\end{acknowledgments}


\bibliographystyle{apsrev4-1}
\bibliography{Literature}

\newpage

\onecolumngrid
\appendix
\newpage
\section{Resonant x-ray characterization of the NNO-NGO SL on NGO (111)$_{\text{pc}}$ with collinear mangetic order }\label{azim}
We use magnetic resonant x-ray scattering to characterize the magnetism in the NNO-NGO SL following the protocol described previously \cite{Frano13, Hepting18}. The experiments were performed at the UE46 PGM-1 end-station at the Helmholtz-Zentrum Berlin using  $\sigma$ and $\pi$ polarized light (perpendicular and parallel to the scattering plane, respectively) and energies tuned to the Ni $L_3$ edge. 
\begin{figure}[h]
\includegraphics[width=0.8\textwidth]{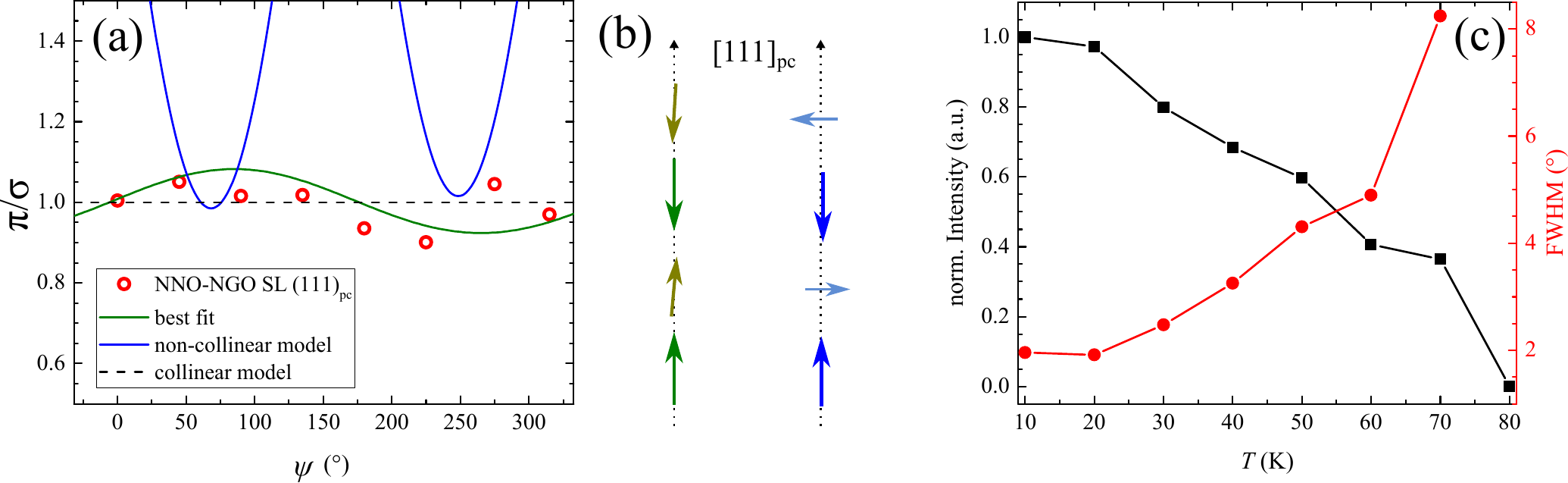}%
\caption{(a) Azimuthal dependence. (b) Sketch of collinear magnetic order (best fit) and spin spiral. (c) Temperature dependence of magnetic REXS signal.  }
\end{figure}

\newpage 
\section{Structural Data} \label{XRD}
\begin{figure}[h]
\includegraphics[width=0.9\textwidth]{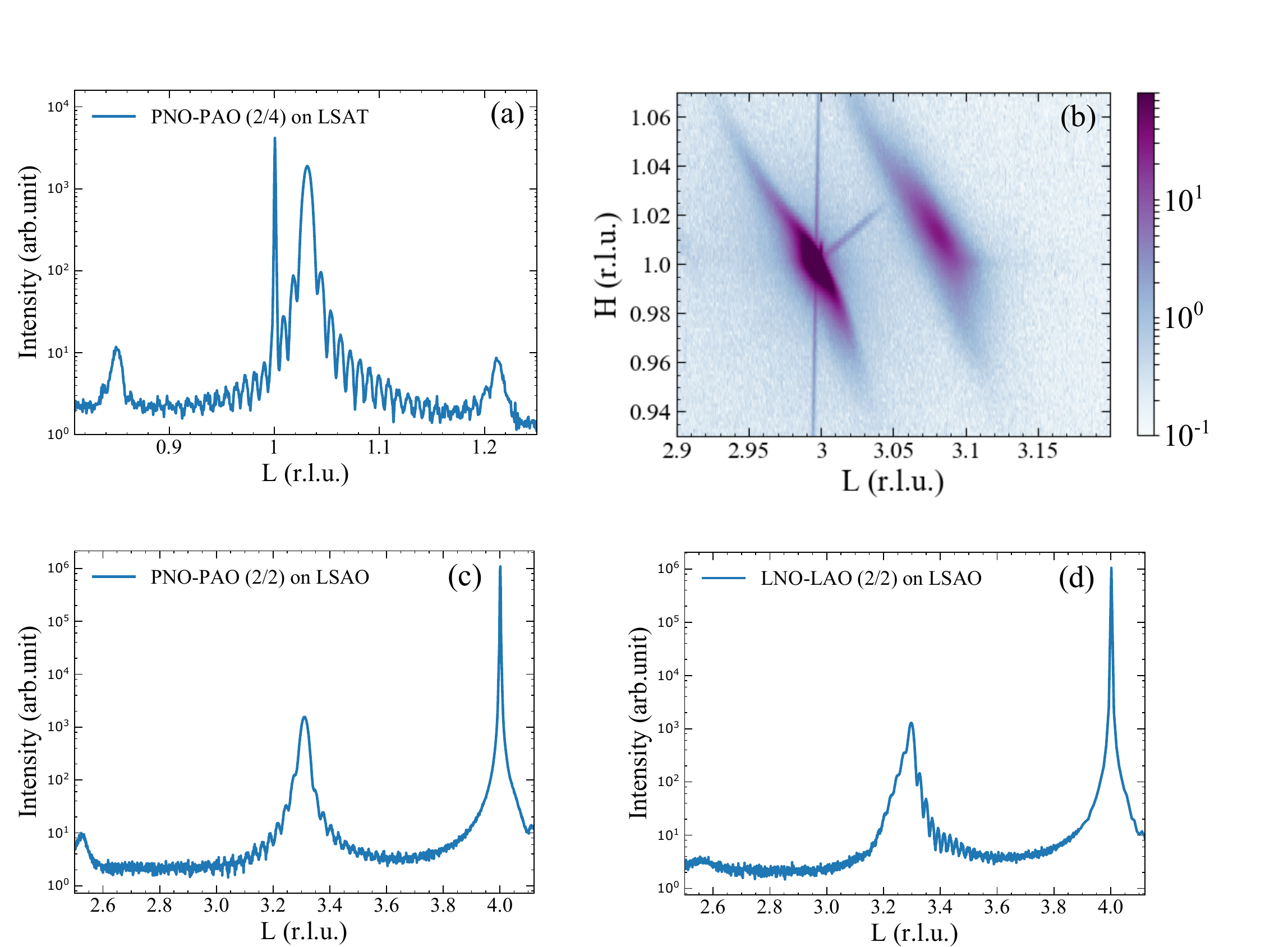}
\caption{(a) and (b): PNO-PAO SL with 2/4 stacking grown on LSAT. (a) L-scan around the (001) reflection. (b) reciprocal space map around (103). The SL is partially relaxed. (c) PNO-PAO SL with 2/2 stacking grown on LSAO. L-scan around (004). (d) LNO-LAO SL with 2/2 stacking grown on LSAO. L-scan around (004).}
\label{XRD_plot}
\end{figure}

Fig.\,\ref{XRD_plot} shows hard x-ray diffraction data of the investigated SLs. The L scans can be seen in panel (a), (c) and (d). We show a representative reciprocal space map around the (103) reflection for the PNO-PAO SL with 2/4 stacking grown on LSAT to illustrate the partial relaxation of the rather thick samples. The most intense peak in the reciprocal space map is from the LSAT substrate. The structural characterization of the NNO thin film can be found in Ref.\,\onlinecite{Lu18}.


\end{document}